\documentstyle[12pt]{article}

\newcommand{\beq}{\begin{equation}}
\newcommand{\eeq}{\end{equation}}
\newcommand{\beqr}{\begin{eqnarray}}
\newcommand{\eeqr}{\end{eqnarray}}
\newcommand{\bitem}{\begin{itemize}}
\newcommand{\eitem}{\end{itemize}}
\newcommand{\p}{\partial}

\renewcommand{\a}{\alpha}
\renewcommand{\b}{\beta}
\newcommand{\g}{\gamma}

\renewcommand{\d}{\delta}
\newcommand{\ep}{\epsilon}

\renewcommand{\l}{\lambda}
\newcommand{\m}{\mu}

\newcommand{\s}{\sigma}

\newcommand{\om}{\omega}

\newcommand{\um}{\frac{1}{2}}
\newcommand{\lag}{{\cal L}}

\begin{document}

\begin{titlepage}

\flushright{FTUV/95-40\\IFIC/95-42\\arch-ive/9507122}
\vspace{1.0cm}
\begin{center}
{\Large{\bf The Role of Temperature in a Dimensional
Approach to  $QCD_3 \,\,\, ^*$}}\\

\vspace{1.0cm}

{\large A.\ Ferrando$^1$ and A.\ Jaramillo$^2$}\\

\vspace{0.2cm}
{Departament de F\'{\i}sica Te\`{o}rica and I.F.I.C.}\\
{Centre Mixt Universitat de Val\`{e}ncia -- C.S.I.C.}\\
{E-46100 Burjassot (Val\`{e}ncia), Spain.}

\vspace{1.0cm}
{\bf Abstract}
\vspace{0.2cm}
\begin{quotation}
{\small\noindent We analyze the role played by temperature in $QCD_3$
by means of a dimensional interpolating approach. Pure gauge $QCD_3$
is defined on a strip of finite width $L$, which acts as an interpolating
parameter between two and three dimensions.
A two-dimensional effective
theory can be constructed for small enough widths
giving the same longitudinal physics as $QCD_3$.
Explicit calculations of T-dependent
$QCD_3$ observables can thus be performed.
The generation of a
deconfinig phase transition, absent in $QCD_2$, is proven
through an exact calculation of the electric or {\em Debye} mass
at high T. Low and high T behaviors of relevant thermodynamic
functions are also worked out. An accurate estimate of the
critical temperature is given and its evolution with $L$ is
studied in detail.}

\end{quotation}

\end{center}
\vspace{1.0cm}
\flushleft{
$^*$Supported in part by CICYT grant \# AEN93-0234 and DGICYT
grant \# PB91-0119. \\
$^1$ ferrando@evalvx.ific.uv.es\\
$^2$ alfonso@goya.ific.uv.es\\
}
\end{titlepage}

\section{Introduction}

In the non-perturbative nature of non-abelian gauge theories
in three and four dimensions stems the complexity of
the confinement phenomenon. On the contrary confinement
shows up as
an outgrowth of one dimensional space kinematics
in  2D non-abelian gauge theories. This is not a striking fact if one
considers the importance of kinematical constraints in 2D
Quantum Field Theories, e.g. in conformal field theories.
It is therefore far from obvious what kind of connection links these
confinement mechanisms in different dimensions, not even
if such connection exists.

In a previous work \cite{fj95} we have studied pure gauge $QCD_3$
on a strip of width $L$. This theory is defined  on a 3D (2+1) space-time
manifold
having pure gauge $QCD_2$ ($L \rightarrow 0$) and
$QCD_3$ ($L \rightarrow \infty$) as extreme limits. An important
property of this dimensional approach is that it allows us to go beyond
the pure $QCD_2$ results in a controlled way (the $\ep$ expansion,
$\ep$ being the dimensionless parameter $\ep=g_3 L^{1/2}$). No matter
how small $\ep$ is, one is dealing with a real 3D theory provided
$\ep \neq 0$. This fact liberates the theory from the strong 2D constraints
we previously mentioned, and at the same time
provides us with a panoply of new physical phenomena.
The explicit solution of the problem is obtained
by the analysis of a 2D effective
theory equivalent to the
original 3D one in the regime of very small widths:
the theory of heavy adjoint matter coupled to 2D gluons
in a gauge invariant way. This theory possesses a
complete glueball spectrum
which can be exactly
computed for small values of the strip width.
Interesting results of various versions of this
model can be found in recent literature \cite{AdjointMatter}.

In the dimensional approach the singularity
and simplicity of the theory's line limit ($QCD_2$)
arises due the absolute dominance of pure longitudinal gluons (2D gluons)
when the width vanishes.
All transverse effects are excluded (brute force!)
by the requirement of pure
2D kinematics. As soon as we allow the strip to acquire a width,
transverse gluons start to play a role.
Their appearance transforms the physical content of the
theory in  a highly non trivial way.
They are responsible for a rich new glueball
spectrum having relevant consequences
in the thermodynamics of the theory.
As it will be shown throughout this paper,
the new physics associated to transverse
gluons is behind the existence of a
deconfining phase transition at high temperature.

In order to understand the connection between the transverse dynamics
and temperature, we will calculate
the free energy of two external static charges in pure gauge $QCD_2$.

 Let us begin by defining our fields in a 2D space-time
with a compactified Euclidean time direction (as usual in
the imaginary time formalism \cite{Kapusta})

\beq
a_\a (x,t) =  a_\a^0 (x) +
\sum_{n\neq 0} a_\a^n (x)\, e^{i2\pi nt/\b }
\label{Fourier_a}
\eeq

\noindent We will work in the $a_1=0$ gauge, where the gluon self couplings
disappear.
The charges are
represented as external static color sources, which we will take
in the fundamental representation and forming a color singlet
state (they behave like heavy quarks in a meson).

Since the sources are static only the gauge field zero mode
will couple to them. Thus we just need to perform
a Gaussian integral to obtain the
free energy of the pair:

\begin{eqnarray}
\lefteqn{\exp \left[ -\frac{1}{T} F_{Q \bar{Q}}(R) \right] = } \nonumber \\
        & & \int [da_0][da_n]\,\,
\exp \left\{ \frac{1}{T} \! \int \! \! dx
\left[- \um a_0^b (-\p_x^2) a_0^b
- \um \sum_{n\neq 0}
        a_n^b (-\p_x^2) a_n^b + i a_0^b j_0^b \right] \right\}  = \nonumber \\
        &=& \exp \left\{ - \frac{1}{T} \int \! dx \, \left[
        \um j_0^b(x) \frac{1}{-\p_x^2} j_0^b(x) \right] \right\}=\exp
[-\frac{\s}{T}
R ]
\label{FreeEnergy}
\end{eqnarray}

The external source corresponding to two static color charges
in the singlet channel is given by $j_0^b=g_2 [Q^b \d (x-x_Q)
+ \bar{Q}^b \d (x-x_{\bar{Q}}) ]$
where $-Q^b \bar{Q}^b = C_N=(N^2-1)/2N$.

Therefore
we obtain a linear growth of the energy with $R$, the string
tension being $\s =  (g_2^2/2) C_N$.
Besides there is no temperature dependence in the free energy.
This is an interesting feature that
is worth analyzing in the framework of the strip formalism.
In a strip with no width, in which there are no transverse
degrees of freedom, the stringlike configuration joining both
charges is the only one contributing to the free energy.
The stringlike configuration is nothing but the classical solution
of static pure gauge $QCD_2$.
Longitudinal perturbations around this classical solution are
forbidden because of the presence of the two infinitely heavy sources.
Transverse perturbations are not allowed by the kinematical constraint
induced by the absence of a width. Therefore, as it is manifest
in the free energy calculation (\ref{FreeEnergy}), the system cannot
jump to any other configuration at any temperature. The probability
of the system to be in the classical stringlike configuration is always
one. In other words the entropy of the system is zero for all temperatures
and thus the free energy becomes T-independent $S=- (\p F/\p T)_V=0$.
Dynamics is frozen and the system lies always in a confining phase.

The previous thermodynamic argument shows already why pure longitudinal
gluon physics cannot ``unfreeze" the 2D string joining the static quarks
in the meson. It is clear now
the relevance of transverse degrees of freedom in this picture.
They are responsible of the thermodynamics of the quark-antiquark
pair.
The possibility of giving some small degree of
transverse dynamics to the system in a controlled way
becomes then crucial to understand the role
of temperature in the previous scenario.
This is precisely the purpose of our contribution.
By using an effective 2D theory
valid for narrow strips
we will try to unveil the interesting
connection between traversal dynamics, confinement
and temperature in pure gauge $QCD_3$.

\section{Characterization of confinement}

The v.e.v. of the Polyakov loop, defined as,

\beq
 <\!0|\mbox{tr}_c g(0)|0\!>_T
\label{PolyakovLoop}
\eeq

\noindent
where

\beq
 g(x) \equiv \exp[i(g_2/T) a_0^a(x) T^a]
\label{PolyakovLoop2}
\eeq
\noindent
 is an order parameter of confinement
depending on temperature. This is so because it is related to
the free energy of an isolated  static color pointlike charge through
the expression \cite{Polyakov},

\beq
 e^{-(1/T) F_Q (T_Q;T)} =  <\!0|\mbox{tr}_c g(0)|0\!>_T
\label{FreeEnergyQ}
\eeq
\noindent
where $T_Q$ is the color isospin of the static charge (we refer to $SU(2)$
-in general we would need
the set of Casimirs characterizing the
representation of $SU(N)$-),
and the matrix $g$ is taken in the representation labeled by $T_Q$.
If the v.e.v. of the Polyakov is zero we are in a confining regime. The energy
to bring the isolated charge from infinity to $x=0$ would be infinite. On the
contrary
if it is not zero, we might liberate it (sending it to infinity) by
supplying a finite amount of energy to the system: deconfinement
occurs! The conventional wisdom
about this order parameter relates it to the discrete symmetry
associated to the center of the gauge group ($Z_N$). Since under
gauge transformations the Polyakov
operator $\mbox{tr}_c g$ is color invariant up to an element of this discrete
group $\mbox{tr}_c g \! \rightarrow \! z_N \mbox{tr}_c g$, its v.e.v.
is also an order parameter of the $Z_N$ symmetry.

Nevertheless we are not going to adopt this approach to characterize
confinement in our problem. Since we describe heavy sources with
quark quantum numbers, the operator $g(x)$ is a $SU(N)$ matrix in
the fundamental representation. As a $SU(N)$ matrix, one can perform
not only local vector transformation upon it, but more general ones.
In particular, we can distinguish left and right color indices and rotate
them independently with $SU(N)$ matrices. In other words,
we can perform $SU(N)_L \otimes SU(N)_R$ transformations on
$g(x)$ in the following way,

\begin{eqnarray}
 g(x) & \rightarrow & U_L(x) g(x) U_R^\dagger(x) \nonumber \\
        U_L(x) & = & e^{i T^a \a_L^a(x)} \nonumber \\
        U_R(x) & = & e^{i T^a \a_R^a(x)}
\label{LRTransformation}
\end{eqnarray}

If we choose $\a_L^a=\a_R^a$ then $U_L=U_R=U$ and we recover the
usual color vector transformations under which
($\mbox{tr}_c g$) is invariant (up to the center). However under
{\em axial} color rotations, defined through the conditions
$\a_L^a=-\a_R^a=\b^a$ ($U_L=U_R^\dagger=U_A$) ,
the Polyakov operator transforms non-trivially and
its trace is no longer invariant
($\mbox{tr}_c g \! \rightarrow \! \mbox{tr}_c (U_A^2 g)$).
Consequently, the v.e.v. of the Polyakov operator becomes
also an order parameter of color axial symmetry
breaking. Ordinary symmetry breaking analysis holds.
If the vacuum preserves the color axial symmetry then
$<\!\mbox{tr}_c g\!>_T=0$ and thus confinement occurs.
If not, this v.e.v. will acquire a non-zero value and the system
will be in a deconfining phase.

We investigate the confinement phase structure of pure gauge
$QCD_3$ on the strip through the analysis of its behavior
under such symmetry. The fundamental field in the study
of the Polyakov loop is the gauge field zero mode. We are
concerned with its transformation rule under $U_A(N)$.
For $SU(2)$ it takes a specially simple form. If $\pi(x)$
stands for the modulus of the gauge field zero mode
and $\b$ for that of the axial angle, then
the referred transformation reads,

\beq
 \pi(x) \stackrel{U_A}{\rightarrow} \pi(x) + 2 \b
\label{AxialTransformation}
\eeq

Color vector transformations keep the modulus unchanged.
Axial ones shift it by a constant. Moreover if we perform
a global vector rotation in such a way the gauge field
zero mode at $x=0$ gets aligned along the 3rd direction
of color isospin, the above transformation is equivalent to
a shift in $a_3(0)$, (from now on  $a_a$ stands for
the zero mode field omitting the $0$-
subscript)

\beq
 a_3(0) \stackrel{U_3}{\rightarrow} a_3(0) + 2 \b
\label{U3Transformation}
\eeq

This is possible because both the strip action
and the operator $\mbox{tr}_c g$ are color singlets.
Due to translation invariance, the v.e.v. of $\mbox{tr}_c g$
is independent of position. Thus Eq.(\ref{U3Transformation})
would apply to any value of $x$. Notice that these transformations
constitute the maximal abelian subgroup of the axial color
non-abelian group $U_A$.

We have found that we can characterize
confinement by studying the behavior of the gauge field zero mode
dynamics under global $U_3$ rotations, as
defined by Eq.(\ref{U3Transformation}).
The vacuum properties of the $a$-field effective theory
become crucial to grasp the
confinement mechanism.

The symmetry
properties of the vacuum can be more easily understood
in the Schr\"odinger formalism.
To study the vacuum structure we need
to know the behavior of the $a$ field
at low energies (long distances).
The long distance action for the gauge field zero mode
will be ruled by the terms with less number of
derivatives,

\beq
 S[a^a] = \int \! dx \left[\frac{1}{2}(1+ Z(a^2)) (\p_x a^a)^2
                        + V(a^2) + ... \right]
\label{LongDistanceAction}
\eeq

\noindent
where we choose $V(0)=0$.
Notice that $Z$ and $V$ depend only on the modulus
of the gauge field due to color invariance.

The previous Euclidean action is one dimensional. Thus we are really dealing
with
a Quantum Mechanical (QM) problem in the Euclidean
``time" $x$. If we focus in the Schr\"odinger
equation for the ground state of the QM counterpart of
Eq.(\ref{LongDistanceAction}), we will find ($a^a \! \rightarrow \! q^a$,
$\p_x a^a \! \rightarrow \! \frac{1}{1+Z} p^a $, $p^a \!= - i \frac{\p}{\p
q^a}$)

\beq
 \left[ -\frac{1}{2 (1+Z(q^2))} \nabla_q^2 + V(q^2) \right] \Psi_0[q^a] =
                                 E_0  \Psi_0[q^a]
\label{GroundStateEq}
\eeq

The vacuum energy $E_0$ is the minimum of all the eigenvalues of the
QM Hamiltonian operator of the above equation. The Euclidean action
(\ref{LongDistanceAction}) is positive defined, thus $E_0 \! \geq \! 0$.
Because of color invariance
the wave functions of the QM problem depend only on the modulus
of the gauge field zero mode $\Psi[q^2]$. Therefore they are
invariant under vector rotations, $q \! \rightarrow \!
U_V q U_V^\dagger$ (recall
the gauge field zero mode is no longer a gauge
degree of freedom). We are free to choose a vector transformation $U_V$
rotating $q$ in such a way the color isospin vector lies always along
the third component axis. That is, $\Psi[q^a]=\Psi[(q^3)_{U_V}]$.
Let us see what happens if besides we demand the
wave function to be invariant under $U_3$ transformations.
This would be precisely the case of the vacuum wave function of the
strip action in the confining phase, as we have seen above.
A $U_3$ symmetric wave function must satisfy,

\beq
\frac{\p}{\p q^3} \Psi[q^3] = 0 \Rightarrow \Psi =
\mbox{{\footnotesize {\it const}}}
\label{U3Invariance}
\eeq

Therefore
the wave function of the $U_3$ invariant vacuum is
independent of the gauge field zero mode.
If we plug the above constant vacuum wave function in
the ground state equation Eq.(\ref{GroundStateEq}),
we find two new properties. First, a $q$ dependent
potential term
is not compatible with such a solution. Thus $V=0$
(recall we have chosen $V(0)=0$).
Second, the ground state energy is zero.

The  $U_3$ invariant vacuum wave
function (confining phase) is a constant whose
energy eigenvalue is zero. Moreover the
effective potential $V$ cannot exist in this phase.
The value of this
constant defining the ground
state wave function, Eq.(\ref{U3Invariance}), is completely arbitrary and
so the wave function is. Thus the ground state
becomes degenerate. This degeneracy is nothing
but that induced by the $U_3$ symmetry.

Let us go into more detail at this point.
A constant gauge field zero mode
($a = \mbox{{\footnotesize {\it const}}}$)
minimizes the action
(\ref{LongDistanceAction}) where there
is no potential $V$ for symmetry requirements.
Therefore each
ground state wave function
is generated by a {\em constant} field $a$ configuration,
and vice versa. We can move from one zero energy configuration to
another one just by changing the value of the constant
field. This is certainly what a $U_3$ transformation does
when acting on a constant $a$ field\footnote{Notice
that for a constant field a $U_3$ transformation is completely
equivalent to a whole axial one $U_A$. We can always choose the
gauge field to be aligned along the 3rd direction for all x.}.

However if the theory develops a non-zero effective
potential for the gauge field zero mode $a$, the
previous property is no longer true.
The presence of an effective potential prevents the
vacuum wave function to be a constant. The $U_3$ degeneracy
is lost. In the broken phase the effective
potential must be different from zero. Since spontaneous breaking
of the $U_3$ symmetry cannot occur in
such a low dimension ($a$
dynamics is defined in $D=1$) the only possibility
for the vacuum to break it is through an explicit
potential term (again this can be seen by
studying the Schr\"odinger equation for
the vacuum wave function Eq.(\ref{GroundStateEq})).

Therefore one can use the effective potential for the
gauge field zero mode itself
as an order parameter of confinement. A zero effective
potential implies we are in a confining phase. A non-zero
effective potential indicates deconfinement occurs.

We can check the previous pattern of confinement in
pure gauge $QCD_2$ Eq.(\ref{FreeEnergy}). In this
simple example the effective action for the $a_0$ is
automatically obtained just by removing the non-zero
mode fields $a_n$. They do not couple either to
the source nor to the zero mode, so they can be
integrated out changing only the
overall normalization constant. After integration
the effective action for the gauge
field zero mode contains only the kinetical
term, which is certainly invariant under
$U_3$ shifts, and no
effective potential appears at any temperature. This agrees
with the fact that pure gauge $QCD_2$ is always
in a confining phase.

\section{The generation of the deconfining phase}

It has been proven \cite{fj95} that a 2D effective action
can be constructed to describe the physics of pure gauge
$QCD_3$ on a very narrow strip. The effective action
corresponds to adjoint scalar Chromodynamics, that is,
to $QCD_2$ coupled to color adjoint matter,

\beq
\lag_{\mbox{strip}} = \mbox{Tr}[ \um f_{\alpha\beta}^2
               +  (D_\alpha \phi)^2 + \m^2_\phi(\ep) \phi^2]
\label{EffectiveAction}
\eeq

\noindent
$\ep$ being the dimensionless parameter $\ep=g_2 L$. The
2D coupling constant $g_2$ is related in turn to the 3D one,
$g_2 \! = \! g_3^2/\ep$.
In the regime where the above action is well-founded, $\ep \ll 1$,
adjoint matter is heavy since

\beq
\m_\phi^2 = \frac{g_2^2 N}{2\pi} \ln(1/\ep^2)
        = \frac{g_3^4 N}{2\pi \ep} \ln(1/\ep^2)
        \stackrel{\ep \rightarrow 0}{\rightarrow} \infty
\label{PhiMass}
\eeq

Eventually it decouples, leaving pure gauge $QCD_2$ as the
effective 2D theory of the widthless strip. Notice that
Eq.(\ref{EffectiveAction}) entitles us to go beyond $QCD_2$
dynamics in a well defined way. The $\phi$ field is nothing
but the zero mode of the transverse component of the
gluon field ($\phi \sim \int_0^L d\!x_2 A_2$). It represents
then a 3D transverse gluon moving on a very narrow strip.
The new dynamics introduced by this transverse degree of
freedom changes radically that of pure gauge $QCD_2$.
3D transverse gluons (in the
effective form of 2D heavy scalar adjoint matter) bind
themselves in color singlets
by means of the confining interaction supplied by the
presence of longitudinal gluons. They make up the
glueball spectrum of $QCD_3$ on a narrow strip
\cite{fj95}.

Apart of being interesting on
its own, the existence of a non-trivial
glueball spectrum has other consequences.
At the very moment the system acquires new configurations
to access to, thermodynamics ceases to be trivial. The partition
function and all the thermodynamical properties derived
from it become
temperature dependent.

In a 3D language, the action (\ref{EffectiveAction})
reproduces the dynamics of transverse fluctuations
around the classical, and unique, solution for $\ep=0$
($A_\m=\d_{\m 0} a_0^{cl} + \phi$). Unlike in
pure gauge $QCD_2$ one may now
excite kinematically the classical string configuration
in the transverse direction by
giving some energy to the system.
Whether these thermodynamical fluctuations
are able to break the longitudinal string confining the charges, or
not, is a matter of the particular dynamics enclosed in the
effective action (\ref{EffectiveAction}).

In order to investigate the high temperature properties
of $QCD$ on the strip, we choose the $a_1=0$ gauge
and select both the gauge and the scalar
field to be periodic in the Euclidean time coordinate.
We can expand them in terms of their 1D Fourier modes
(as in Eq.(\ref{Fourier_a})) and obtain the following 1D
strip action at temperature $T$,

\beq
S(T) = \frac{1}{T} \int \! \! dx \,\, \left\{
        \um \phi_n^a \left[ (- \p_x^2 + \mu^2(\ep) + \om_n^2) \d_{n,-n'}
\d_{ab}
        + (M_1)_{ab}^{n,-n'} + (M_2)_{ab}^{n,-n'} \right] \phi_{n'}^b \right\}
\label{ActionT}
\eeq

\noindent
where,

\begin{eqnarray}
  \om_n  & \equiv  & 2 \pi n T \nonumber \\
  (M_1)_{ab}^{n,-n'} & \equiv & - 2 i g_2 \ep^{abc} \om_n a^c_{-n'-n} \nonumber
\\
  (M_2)_{ab}^{n,-n'} & \equiv &  g_2^2 \ep^{abc} \ep^{bdf} a^c_{n''}
a^c_{-n-n'-n''}
\label{ActionTDefinitions}
\end{eqnarray}

The $\phi$ integration can be formally done yielding the
effective action for the gauge field modes,

\beq
 S_{\mbox{{\footnotesize{\it eff}}}}[a_n] = \frac{1}{T} \int \!  \! dp \,\,
        \um \mbox{Tr} \left[ \ln (p^2 + \mu^2 + \om^2 + M_1 + M_2) \right]
\label{ActionTModes}
\eeq

\noindent
Now the trace runs over both color and the mode index $n$.
The calculation of that trace is impossible in the most general case.
However expression (\ref{ActionTModes}) is useful to determine
the limits in which a well defined approximation can be
established.

We are concerned with the possibility that
the theory undergoes a deconfining transition at
high temperature. According to what was discussed
in the last section, we need to know whether the $\phi$ dynamics
is able to generate a non-zero effective potential for $a_0$
or not. At zero $T$ such a term is not allowed since
$a_0$ is a truly gauge degree of freedom in that case.
Local gauge invariance prevents its existence.
At finite $T$ the gauge field zero mode becomes a covariant
object under gauge transformations. As a consequence we cannot
longer resort to gauge invariance to protect the theory from getting
a non-zero effective potential. This possibility becomes therefore a dynamical
issue.

At very high $T$ and for very narrow strips, for which adjoint
matter is very heavy, the following inequalities  hold,

\begin{eqnarray}
        T \gg  \mu(\ep) & \gg & g_2  \nonumber \\
         \om_{n \neq 0}^2 + \mu^2(\ep) & \gg  & \mu^2(\ep) \gg g_2^2
\label{Inequalities}
\end{eqnarray}

The previous regime legitimates a perturbative expansion of the
logarithm in the effective action in terms of
the gauge field modes (\ref{ActionTModes}). The
coupling constant dependence arises through the $M_1$ and
$M_2$ operators Eq.(\ref{ActionTDefinitions}).
Their contributions are small when compared to those arising
from the non-interaction terms.

After performing the gauge field expansion keeping only
terms up to order $g_2^2$, one obtains a typical finite $T$
1-loop calculation for the mass of the $l$-mode (see Fig.1),

\beq
 m_l^2(\ep,T) = g_2^2 \, T \left\{
        \sum_n \! \int \! \frac{dp}{2 \pi} \frac{1}{p^2 + \mu^2 + \om_n^2}
        - \sum_n \! \int \! \frac{dp}{2 \pi}
        \frac{\om_n^2+ \om_n \om_{l+n}}
        {(p^2 + \mu^2 + \om_n^2)( p^2 + \mu^2 + \om_{l+n}^2)} \right\}
\label{Mass}
\eeq

The first, and very important, property of the above mass is that it is zero
for all non-zero modes. Every non-zero mode is massless to this
order of the approximation. We will point out
the relevance of this feature in our conclusions.

On the contrary, the gauge field zero mode acquires a non-zero
mass which is computed systematically under the conditions
given by the inequalities in Eq.(\ref{Inequalities}). Therefore at
high temperatures the electric, or Debye, mass is given by

\begin{eqnarray}
 m_{el}^2(\ep,T) & = &  g_2^2  \left\{ \frac{T}{2 \mu} - \frac{1}{2 \pi}
        + \frac{\mu^2}{8 \pi^3 T^2} \zeta (3) -
        \frac{3 \mu^4}{64 \pi^5 T^4} \zeta (5) +
        \frac{15 \mu^6}{1024 \pi^7 T^6} \zeta (7) \right\} \nonumber \\
         & &  \nonumber \\
         & & + \mbox{{\cal O}} [(\frac{\mu}{T})^8, (\frac{g_2}{\mu})^4)]
\label{ElectricMass}
\end{eqnarray}

Moreover one can prove that, within the regime determined by the conditions
(\ref{Inequalities}), there are no corrections to the derivative
dependent part of the effective action (\ref{ActionTModes})
(they are suppressed as powers of $g_2^2/\mu^2$). In this way
we get an exact high temperature result for the zero mode effective
action of $QCD$ on the strip,

\beq
 S_{\mbox{{\footnotesize{\it eff}}}}[a_0] =  \frac{1}{T} \! \int \!\! dx
        \left[\um a_0^b (-\p_x^2) a_0^b  + \um m_{el}^2(\ep,T) a_0^b a_0^b
\right] +
        \mbox{{\cal O}} (\frac{g_2}{\mu})^4
\label{EffectiveActionZeroMode}
\eeq

Our question about whether transverse gluon fluctuations, even
if small,
would be able to break the tube flux structure at high enough temperature
has now a clear answer. According to our discussion in the
previous section the appearance of the mass term in
Eq.(\ref{EffectiveActionZeroMode}) is enough to ensure the
breaking of the $U_3$ axial symmetry and, as a consequence,
the generation of a non-vanishing confinement order parameter
($<\!\mbox{tr}_c g\!> \neq 0$). We conclude then
that {\em $QCD_3$ defined on a strip experiences a deconfinig
phase transition at high temperature}.

But we can go further, for we can explicitly
check this result by calculating the Wilson loop ($W(T,R)$) in an
analogous way as we did in Eq.(\ref{FreeEnergy}).
We only have to substitute the linear 1D propagator
of pure gauge $QCD_2$ by the exponentially decreasing
one derived from the action (\ref{EffectiveActionZeroMode}).
The functional integration is immediate because is quadratic
in $a_0$,

\begin{eqnarray}
-\ln W(T,R) & = & \frac{1}{T} F_{Q\bar{Q}}(T,R)
         \stackrel{T \gg \mu \gg g_2}{\approx} \frac{1}{T} \int \! dx \,
        \um j_0^b(x) \left( \frac{1}{-\p_x^2 + m_{el}^2} \right) j_0^b(x)
\nonumber \\
        &=& \frac{1}{T} \frac{\s}{m_{el}}(1- e^{- m_{el} R})
                \stackrel{R \gg g_2^{-1}}{\rightarrow} \frac{\s}{m_{el} T}
\label{FreeEnergyDeconfinement}
\end{eqnarray}

In the $a_1=0$ gauge the Wilson loop can be easily related to the
Polyakov operator, $W(T,R) \!=\! <\!\mbox{tr}_c[g(R)g^\dagger(0)]\!>$.
Because of cluster decomposition, our above result certainly implies,

\begin{eqnarray}
W(T,R) \stackrel{R \gg g_2^{-1}}{\rightarrow} <\!\mbox{tr}_c g(0)\!>^2
        & \Rightarrow & <\!\mbox{tr}_c g \!>
                = e^{- \frac{1}{2 T} \frac{\s}{ m_{el}} } \neq 0
\label{Cluster}
\end{eqnarray}

In agreement with the general arguments given in the previous section.
Notice how the absence of a mass is the responsible of the vanishing
of the v.e.v. of the Polyakov operator. The strong IR behavior of
the gauge field zero mode turns the ordinary analytical expansion
of the exponential operator into an essential singularity.

\section{Low temperature glueballs}

The theory at zero temperature exhibits a confining behavior.
The Wilson loop fulfills an area law at long distances in the
same way as pure gauge $QCD_2$. The only difference
appears in the exact value of the string tension, which is
now renormalized by quantum transverse fluctuations
($\phi$-loops) \cite{fe95,ht95}. The spectrum of
the theory at small $\ep$ can be found in reference
\cite{fj95}. The low lying spectrum is calculated
explicitly and is formed by heavy glueballs, made
up of two heavy constituent transverse
gluons. The radial wave functions are proportional to
Airy functions with energies given by,

\begin{eqnarray}
 M_r(\epsilon) & =  & 2\mu_R(\ep) +
  \varepsilon_r \left(\frac{\sigma^2}{\mu_R(\ep)}\right)^{1/3}
\label{GlueballMass}
\end{eqnarray}

($-\varepsilon_r$ are the zeroes of the Airy function and $\s$ is
the $\phi-\phi$ string tension). The {\em renormalized}  scalar mass
$\mu_R$ is obtained by using a self-consistent treatment
of the IR divergences \cite{dh82},

\beq
 \mu_R^2(\epsilon) = \frac{g_2^2 N}{4 \pi} \ln(\frac{1}{\epsilon^2})
\label{RenormalizedscalarMass}
\eeq

The low temperature behavior of the partition function
will be ruled by the lowest mass states of the
physical Fock space. They are eigenstates of the
Hamiltonian operator at zero temperature. Because
confinement occurs at zero temperature the only
physical states contributing to the partition function
are  glueballs (singlet $\phi-\phi$ boundstates).
In the most general case (even when vacuum
excitations and non-planarity effects are suppressed,
e.g. in the $N \! \rightarrow \! \infty$) the eigenstates
of the full Hamiltonian will not coincide with those
obtained by solving the bound state equation of the
$\phi-\phi$ system Eq.(\ref{GlueballMass}). This is so
because glueballs can
interact among them in an effective way \cite{Princeton} and a
further mass renormalization can
occur.

In our case we are going to consider glueball states
as almost-free particles, exposed only to contact interactions.
Therefore our Hamiltonian eigenstates will be given by the
solutions of the two body $\phi-\phi$ bound state equation.
This approximation is completely justified
for small enough $\ep$ since for very heavy glueballs, as it is the
case, the glueball interaction can be neglected. The OBEP (One Boson Exchange
Potential) is extremely short range ($\sim 1/M_{GB} \ll R_{GB}$ ,
$M_{GB}$ being the glueball mass and $R_{GB}$
the glueball radius), as one can easily estimate using the strip results.
The glueball radius can be computed using the known radial wave
functions. If we call $R_0 \equiv  (\mu_R \sigma)^{-1/3}$ then the glueball
radius can be
obtained from the {\em mean squared radius} of the two
particle system,

\beq
 R_{GB} \equiv  <\!x^2 \!>^{1/2} \simeq 1.707510 \; R_0
\label{GlueballRadius}
\eeq

Therefore in the small $\ep$ limit, the correlation length of
the glueball exchange is negligible when compared to
the typical radius of the glueball because

\beq
\frac{\xi_{GB}}{R_{GB}} \simeq 1/(M_{GB} R_0)
                                \stackrel{\ep \ll 1}
                                {\simeq}(\frac{\sigma}{\mu_R^2})^{1/3}
                                         \stackrel{\ep \rightarrow 0}
                                                        {\longrightarrow} 0
\label{CorrelationLength}
\eeq

On the other hand the typical radius
of a glueball $R_{GB}$ decreases logarithmically as the strip
width tends to zero,
        $R_{GB} \! \sim \! R_0 \! = \! (\mu_R \sigma)^{-1/3} \! \sim \!
                (\ln(1/\ep^2))^{-1/3} \!
                \stackrel{\ep \rightarrow 0} {\longrightarrow} \! 0$. We
conclude
that glueballs are extended small objects  in the low $\ep$
regime.

Consequently our system is made up of small quasi-free glueballs in the small
width regime. Moreover since they are also very heavy, a non-relativistic
approach
(NR) is likewise appropriate at low temperature. For this reason
the NR free glueball gas model becomes an excellent  approximation to
the real theory at low temperature. Since the dimensionless width $\ep$ is a
free
parameter, we can play with it in order to make
our approach as good as possible. In this way
the narrower the strip is the better the free gas  approximation is.

The thermodynamics of the NR  gas of glueballs is described by a
partition function which factorizes in individual partition functions
for each glueball state. At fixed volume, the pressure ($\equiv T  \ln Z/V$)
is then,

\beq
 P_{GB}(T;\ep) = P_1 (T;\ep) + P_2(T;\ep) + \cdots + P_r(T;\ep) + \cdots
\label{Pressure}
\eeq

 The general term in Eq.(\ref{Pressure}) corresponds to the
pressure of a gas of 2D NR scalar $r$-particles at temperature T with
masses $M_r$ given
by Eq.(\ref{GlueballMass}). That is,

\beq
 P_r(T;\ep) =
\exp \left( -\frac{M_r(\ep)}{T} \right)
        \left( \frac{M_r(\ep)}{2 \pi} \right)^{1/2} T^{3/2}
\label{PressureMr}
\eeq

An expected consequence of the previous result is that low T
pressures vanish when we shrink the width strip
to the line limit (recall $M_r(\ep) \! \stackrel {\ep \rightarrow
0}{\rightarrow}
\! \infty$). However that is not the whole story. We can
go beyond our qualitative expectation and obtain the
small $\ep$ behavior of the pressure exactly. If we make the $\ep$ dependence
in Eq.(\ref{PressureMr}) explicit by using Eqs. (\ref{GlueballMass})
and (\ref{PhiMass}), we find a non-analytic form for its
$\ep \! \rightarrow \! 0$ limit,

\begin{eqnarray}
\ln Z_{GB}/V = P_{GB}(T;\ep)  \approx
        e^{-\g \frac{(- \ln \ep)^{1/2}}{\ep}}
                \,\,\,  (- \ln \ep )^{1/2}
        & \stackrel{\ep \rightarrow 0}{\rightarrow} 0 &
                \g = \frac{g_3^2}{T} (\frac{N}{\pi})^{1/2} \nonumber \\
        & &
\label{Epsilon}
\end{eqnarray}

\section{Near the phase transition}

Pure gauge $QCD$ on a narrow strip turns out to be a two phase
theory. In section 3. we have studied its structure at very high T.
We encountered a deconfining regime in which both transverse
($\phi$-particles) and longitudinal gluons (2D gluons) are the
physical degrees of freedom. In contrast to the high T behavior,
the Fock space of the theory at low T is made up of color
singlet glueballs. Thermodynamics of the ``cold'' phase has
no much to do to that of the free gluon plasma.
Thermodynamic functions of the former
correspond to an admixture of infinite species of glueball gases
characterized by their different masses
Eq.(\ref{Pressure}). In the latter only transverse gluons,
as a $\phi$ field gas, contribute to the partition function.

We are going to investigate the critical behavior of the system both in its
gluon plasma phase and in its glueball phase. If we are able to
establish accurately the thermodynamic properties of the two phases
in the neighborhood of
the phase transition point, we will be in an optimal
position to estimate its critical temperature.

Let us begin analyzing the plasma phase. The starting point is the effective
action for the gauge field modes Eq.(\ref{ActionTModes}),
obtained after $\phi_n$ integration. The partition function
is then just,

\begin{eqnarray}
  Z_{\mbox{{\footnotesize {\it plasma}}}} (T;\ep) & = &
        Z_\phi(T;\ep) \int \! [da_n]
        e^{-S_{\mbox{{\footnotesize{\it eff}}}}[a_n]/T} \nonumber \\
        Z_\phi(T;\ep)  & \equiv  &
                e^{-\frac{1}{T} \int \!  \! dp \sum_n
        \um \mbox{tr}_c \left[ \ln (p^2 + \mu^2 + \om_n^2) \right]}
\label{PartitionFunction}
\end{eqnarray}

Certainly the latter operation is not feasible in the most general case.
However we will not need to work it out in an exact manner. Only
very general properties of the functional integration will be required
to understand the problem under consideration.
The existence of {\em two} independent dimensionless parameters
is the main ingredient to see why a complete knowledge of
the effective action is not necessary. Because the theory
possesses three parameters with dimension of mass
($g_2, \mu,T$), we can form the dimensionless ratios
($ \mu/T, g_2/\mu$). Thus we can express
any relevant function or functional in terms of them.
For example, the high T calculation in section 3. is
specified by the conditions ($\mu/T \! \ll \! 1, g_2/\mu \! \ll \! 1$).
However nothing forces us to choose {\em a priori}
$\,\, \mu \!< \! T$, being both variables independent. We can
take advantage of the freedom to select an arbitrarily large
mass by working on a narrow enough strip.
So that at any given T we can fulfill the conditions
$\mu \! \gg \! g_2, \mu \! \gg \! T$
just by picking a small enough value of $\ep$.
The first inequality is understood in the same way as in the
calculation of the spectrum at zero temperature. That is
the interaction term is small when compared to the kinetic
piece of the bound state equation (NR approach). In this
approach higher order contributions of the gauge field are suppressed
as powers of $g_2/\mu$. Therefore
at any T we expect a weak coupling expansion
of the effective action Eq.(\ref{ActionTModes})
in the gauge field modes to be valid when $\ep \! \ll \! 1$.

The structure of  $S_{\mbox{{\footnotesize{\it eff}}}}[a_n]$ in the weak
coupling
regime is easy to obtain from
Eq.(\ref{ActionTModes}) by expanding the logarithm in powers
of $(p^2 + \mu^2 + \om^2)^{-1}(M_1+M_2)$,

\begin{eqnarray}
 S_{\mbox{{\footnotesize {\it eff}}}}[a_n] & = &
        S_{der}[\p_x a_n] + V_{\mbox{{\footnotesize{\it eff}}}}[a_n]
                                                \nonumber \\
V_{\mbox{{\footnotesize{\it eff}}}}[a_n]  & = &
        \int \! dx \left\{ \um  m_0^2 (a_0^b)^2 +
                \sum_{n_1+n_2+n_3=0}
                        \l_{n_1 n_2 n_3}^{abc} ( a_{n_1}^a a_{n_2}^b a_{n_3}^c)
                                + \cdots  \right\}           \nonumber \\
S_{der}[\p_x a_n] & = &  \int \! dx \left\{
        \um \sum_n (1 + \l'_n) (\p_x a_n^b)^2 + \cdots  \right\}
\label{ActionStructure}
\end{eqnarray}

Although it is impossible to determine an explicit solution for all the
coefficients, it is at our hand to establish their behavior
for small $\ep$.
We were able to calculate the first term of the effective potential
in section 3. Although we only gave its high T expansion, it is
also possible to show its form for $\mu \! \gg \! T$. The frequency
sum in Eq.(\ref{Mass}) can be transformed in an integral over
the zero component of the momentum $p_0$ \cite{Kapusta}.
The T-dependent part of this integral can be worked out,
in the same way as the
remaining $p_1$ integration when we work in the
$\mu \gg T$ regime.  The whole operation
results in a exponentially decaying
behavior for the electric mass,
$\sim \! e^{- \mu/(2 \pi T)} \! \stackrel{\ep  \rightarrow  0}
{\rightarrow} \! 0$.
The exponential decay of the action coefficients
is a general feature valid
at every order for large enough masses. As a consequence all
the coefficients of the different powers of $a_n$ appearing in
Eq.(\ref{ActionStructure}) vanish in the $\ep \! \rightarrow \! 0$
limit. Moreover one finds that coefficients of higher
powers decay more strongly than those of the lower ones.
The above characteristic of the effective action coefficients
is completely understandable within a 3D framework.
We know the 3D gauge theory tends to its 2D counterpart
in the line limit. Thus all coefficients in the effective potential
must disappear  when  $\ep \! \rightarrow \! 0$ leaving
the pure $QCD_2$ action as the only remnant of color interaction.

In summary. We find that at any given T, {\em above} the
critical temperature, we can reduce the contribution from
the effective action to that of a 2D gluon gas provided
we take the strip width in a convenient way. Indeed
the $\ep \! \rightarrow \! 0$ limit is T-independent
and therefore this limit is equivalent to write
simultaneously ($\mu \! \gg \! g_2, \mu \! \gg \! T$)
when T is taken at a fixed value.

Mathematically we can express the above
property as follows,
according to Eqs.(\ref{PartitionFunction})
and (\ref{ActionStructure}),

\begin{eqnarray}
        Z_{\mbox{{\footnotesize {\it plasma}}}} (T;\ep)
        & \stackrel{\ep \ll 1}{\approx} &
                Z_\phi(T;\ep) \,\, Z_{QCD_2}
                                                \nonumber \\
        Z_{QCD_2}  & \equiv & \int \! [da_n]
                e^{-\frac{1}{T} \! \int \!\! dx \sum_n
        \left[\um a_n^b (-\p_x^2) a_n^b \right]}
\label{PartitionFunction2}
\end{eqnarray}

That is, the partition function of the gluon plasma
factorizes in two pieces. One corresponds to a
free gas of transverse gluons ($Z_\phi$) and the
other one to a free gas of longitudinal gluons
($Z_{QCD_2}$). The $QCD_2$ gluon partition
function is trivial. As mentioned in the introduction,
it is T-independent and thus it has no
thermodynamic significance as we will
see in a moment.
All the temperature dependence is produced by the
quasi-free motion of transverse gluons.
The plasma pressure is immediately obtained from
Eq.(\ref{PartitionFunction2}) using standard
finite T field theory techniques,

\beq
 P_{\mbox{{\footnotesize {\it plasma}}}} (T;\ep) \stackrel{\ep \ll 1}{\approx}
        P_\phi (T;\ep) + P_\phi^{\mbox{{\footnotesize {\it vac}}}} =
                (N_c^2 -1 ) e^{-\mu(\ep)/T}
                        \left( \frac{\mu(\ep)}{2 \pi} \right)^{1/2} T^{3/2}
                                + P_\phi^{\mbox{{\footnotesize {\it vac}}}}
\label{PlasmaPressure}
\eeq

Now comes the time to find out the properties of the glueball
phase above the critical temperature.
In section 3. we discovered the way the glueball admixture started to react
under a small increase of the heat bath temperature.
Only the lowest mass glueballs produced a discernible effect on the
gas pressure, as Eq.(\ref{Pressure}) shows. The
contributions to the global pressure of gases made
out of heavy glueballs are exponentially
suppressed at low T. In general, at any T
the pressure of a given class of glueball
is always exponentially suppressed respect to
that of lighter boundstates.

The transverse gluon
$\phi$ is lighter than all its glueball boundstates.
The pressure of the
$\phi$ gas
is always higher than any individual pressure of the glueball soup.
Only the fact that we need to sum up a great
amount of single contributions, from the infinite number of glueball
states we can create from $\phi$ particles, can avoid that
the total pressure of the glueball phase remains below
the plasma pressure at any T.
According to Gibbs criterion, if
$P_{GB} < P_{\mbox{{\footnotesize {\it plasma}}}}$ at any T
then the transition cannot occur.

The mass distribution of the states becomes then crucial. The
particular form of the density of the states determines
the real shape of the pressure and thus the possibility
for the phase transition to happen. One can really
appreciate this point when writing the total pressure as

\beq
 P_{GB} = \int_0^{\infty} \! dm \rho (m) e^{-m/T}
                \left( \frac{m}{2 \pi} \right)^{1/2} T^{3/2}
\label{PressureDensity}
\eeq

If the density of states increases rapidly for
large masses ($m \gg T$) then it can compensate
the thermodynamical suppression dictated by
the Boltzmann factor and produce the necessary
cut. The density of high energy states derived from the
NR spectrum (\ref{GlueballMass}) generates the
following high T-dependence of the glueball pressure,

\beq
 \rho (m) \stackrel{m \gg \mu_R}{\sim} m^{1/2}
        \Rightarrow P_{GB} \sim T^{7/2} >
        P_{\mbox{{\footnotesize {\it plasma}}}}  \sim T^{3/2}
        \,\,\,\, \mbox{{\it at large T}}
\label{TGlueballPressure}
\eeq

As we can see, the transition occurs (see Fig.2b).

However although we can take for granted the existence
of the deconfining transition, we are not describing
properly the glueball phase near $T_c$. The reason is
we cannot
extrapolate the NR spectrum (\ref{GlueballMass})
to the highest energy boundstates. The accuracy of the calculation
is based on the validity of the NR approach.
For very energetical boundstates the binding energy is  much
bigger than $\mu_R$. The typical NR parameter
$V_{\mbox{{\footnotesize {\it int}}}}/\mu_R$,
an estimation of the characteristic relative velocity, is no
longer small and a full relativistic approach has to
be adopted instead. Moreover the heavy part of the
spectrum has to contain necessarily multi-$\phi$
states, not only the two particle sector. Pair excitations
are now energetically available since
typical momenta flowing through internal gluon
lines are much bigger than the pair energy threshold.
This means that our NR counting of glueball states is
clearly underestimating the real number of them
contributing to the total pressure. A careful counting
of states is necessary in order to determine an accurate
description of the glueball gas pressure
near the critical temperature $T_c$.

Like in $QCD_2$ with fermions,
one can use light cone quantization and obtain a full relativistic
Bethe-Salpeter equation for multi-$\phi$ bound states valid
for any constituent mass. This study has been already carried out
quantitatively by {\em Demeterfi et al.} \cite{dk94}.
In the description of the highest spectrum the constituent mass
can be neglected respect to the boundstate mass ($m \! \gg \! \mu_R$).
Thus we can study the growth of the density of the
most massive states by analyzing
their model in the limit in which the constituent mass equals zero.
Their results in this limit are very interesting since they show that the
density
of states exhibits the Hagedorn behavior \cite{ha65},

\beq
 \rho_H (m) \approx g_2^{b-1} a  m^{-b} e^{m/T_0}  \,\,\,\,\, [a]=1
\label{Hagedorn}
\eeq

Although only an estimation of the Hagedorn temperature
$T_0(\ep) \! = \! (1.4 \! - \! 1.5) \sqrt{N_c/\pi } (g_3^2/\ep)$
is given, the previous result is enough to obtain the relevant
contribution to the glueball gas pressure near $T_c$.
The ultrarelativistic boundstates
are much more abundant than the NR ones (compare
the exponential behavior in Eq.(\ref{Hagedorn}) to
the NR density). This results in a stronger raise of
the glueball pressure with T and therefore in a
decrease of the critical temperature (see Fig.2a).
The ultrarelativistic boundstates dominate the region below
the phase transition point, as we will see next.

In the pressure integral
(\ref{PressureDensity}) it is legitimate to substitute the
density of states by $\rho_H$ provided we
work at energies above a typical mass $M = \l \mu$
characterizing the ultrarelativistic regime. That is, $\l \! \gg \! 1$.
All states with masses bigger than M will be well described
by Eq.(\ref{Hagedorn}).

\begin{eqnarray}
 P_{H}(T;\ep) & \stackrel{T \approx T_0 \,\, \ep \ll 1}{\approx} &
                \int_{\l \mu(\ep)}^\infty \! \! dm \,\,
                \rho_H (m) e^{-m/T} \left( \frac{m}{2 \pi} \right)^{1/2}
T^{3/2}
                                + P_{GB}^{\mbox{{\footnotesize {\it vac}}}}
                                        \nonumber \\
                & \sim &  (\mbox{{\it constant}}) \,\, \mu(\ep) ^{1/2-b}
                        e^{-\l \frac{\mu (\ep)}{T_0(\ep)}
(\frac{T_0(\ep)}{T}-1)}
                                (\frac{T_0(\ep)}{T}-1)^{-1}
                                + P_{GB}^{\mbox{{\footnotesize {\it vac}}}}
                                                \nonumber \\
                        & &
\label{PressureHagedorn}
\end{eqnarray}

In principle, the constants $P_\phi^{\mbox{{\footnotesize {\it vac}}}}$
and $P_{GB}^{\mbox{{\footnotesize {\it vac}}}}$ in
Eqs. (\ref{PlasmaPressure}) and (\ref{PressureHagedorn})
may be different. The former corresponds to the pressure of the
non-confining vacuum (free gluons), whereas the latter
corresponds to the confining vacuum generating the Fock space of
singlet color states.
The difference between these two
pressures is given by the difference between the vacuum
thermodynamical potentials of the two phases
($\Omega = -P$)

\beq
 P_\phi^{\mbox{{\footnotesize {\it vac}}}}-
        P_{GB}^{\mbox{{\footnotesize {\it vac}}}}=
\Omega_{\mbox{{\footnotesize  non-conf}}}^{\mbox{{\footnotesize {\it vac}}}}
-\Omega_{\mbox{{\footnotesize conf}}}^{\mbox{{\footnotesize {\it vac}}}} =
- {\cal E}_{\mbox{{\footnotesize conf}}}^{\mbox{{\footnotesize {\it vac}}}}
\label{VacuumPressure}
\eeq

\noindent
where
${\cal E}_{\mbox{{\footnotesize conf}}}^{\mbox{{\footnotesize {\it vac}}}}$
is the vacuum energy of the confining phase.

The vacuum structure of
pure gauge $QCD_2$ is trivial for it is a free massless theory,
as it is apparent in a physical axial gauge (e.g. $a_1 \! = \! 0$).
Obviously it is manifestly invariant under dilatation transformations
and thus its vacuum energy is zero after zero point
renormalization
(${\cal E}_{\mbox{{\footnotesize {\it vac}}}} \! \sim
        < :\!T_{\a\a}\!:>
        \sim <: \! \p_\a j_\a^{\mbox{{\footnotesize dil}}}\!:> = \! 0$).

Pure gauge $QCD_3$ on a narrow strip is a confining theory.
However the vacuum
structure derived from the strip action (\ref{EffectiveAction}) is
the same as in pure gauge $QCD_2$.
After integrating the transverse
fluctuation $\phi$ we obtain an effective action for the 2D gluon field
whose first term is identical to that of pure gauge $QCD_2$. This
term renormalizes the coupling constant but in a {\em momentum
independent} way, $g_2^R \! =   \!  Z(\mu) g_2$. Thus no breaking
of dilatation symmetry is encountered and we meet the same result
as before \cite{Shuryak}

\beq
{\cal E}_{\mbox{{\footnotesize conf}}}^{\mbox{{\footnotesize {\it vac}}}}
        \sim
        < :\!T_{\a\a}^{\mbox{{\footnotesize strip}}} \!:>
        \sim <: \! \p_\a
(j_\a^{\mbox{{\footnotesize dil}}})^{\mbox{{\footnotesize strip}}}\!:> =  0
\label{VacuumEnergy}
\eeq

The previous result can be likewise related to the absence of a gluon
condensate ($<\!f^2 \!> \!=\! 0$) in the theory
represented by the strip action (\ref{EffectiveAction}). As
we mentioned before the long distance effective action
is proportional to that of pure gauge $QCD_2$, which
being a free massless theory cannot generate a non-zero
gluon condensate.

According
to Gibbs criterion, at the critical temperature it must be
true, (see Fig.2)

\beq
        P_{GB} (T_c;\ep) =
        P_{\mbox{{\footnotesize {\it plasma}}}} (T_c;\ep)
\label{Equalization}
\eeq

Once we know the two constants appearing in the
expressions for the gluon and glueball gas pressure
are the same, the equalization
equation yielding $T_c$ is completely defined. The two
constants cancel in Eq.(\ref{Equalization}) just rendering
an equation for the T-dependent parts of Eqs. (\ref{PlasmaPressure})
and (\ref{PressureHagedorn}). We can now
find the evolution of $T_c$ with the width
valid for narrow strips.

{}From the 3D point of view the two dimensional coupling constant
is a width dependent object, $g_2= g_3^2/\ep$. If we wished to
compare to numerical simulations in 3D, the natural variables
to work with would be $(g_3,L)$ or $(g_3,\ep)$. The 3D coupling
constant $g_3$ defines completely the color flux generated by
the static quark-antiquark pair in the infinite volume limit.
The string tension, for example, can be expressed in terms of it since
$g_3^2$ carries dimensions of mass. If we put the system
on a strip (an asymmetric box, in the case of a lattice
simulation), the natural question to answer is
what happens to the infinite volume values of
relevant observables such as the string tension,
the glueball masses or the critical temperature under
the new different boundary conditions. This is
equivalent to fix the value of $g_3$ (maintain the
infinite volume quark-antiquark configuration)
and tune only the strip width $L$.

In Fig.3 we give an approximate representation of the
critical temperature as a function of $\ep$. It is obtained
by solving numerically the equalization equation (\ref{Equalization}) for
different values of $\ep$.
We have observed that the discrete
contribution $P_{GB}$ is not relevant numericaly.
However it is worth observing that the critical temperature does not exist in
the exact
line limit of the strip.
This is not because
the effective 2D coupling constant blows up in that limit,
which pushes $T_c$ to infinity.
In the strict $\ep \! \rightarrow \! 0$ limit both the
plasma and the glueball gas pressure are exactly zero independently
of temperature. The equalization equation has no solution
for $\ep=0$. The previous feature is not related to the dimensional
running of the 2D coupling constant, but to the
effective decoupling of the transverse gluon
$\mu_R \stackrel{\ep \rightarrow 0}{\rightarrow} \infty$
in that limit. Therefore
we recover in our strip approach a previously known result,
pure gauge $QCD_2$ is always in the same (confining)
phase.

{}From the previous analysis one concludes that
the existence of the two phase
structure is a non-trivial quantum effect produced by the interaction
between longitudinal and transverse degrees of freedom.
Simpler models, as the non-compact 3D abelian gauge theory,
lacks such property. As a consequence, they do not show
the rich phase structure of the non-abelian theory \cite{fe95}.

\section{Conclusions}

Our study establishes the preeminent role of transverse dynamics in the
generation of a deconfining phase. Unlike their longitudinal
counterparts, transverse gluons are dynamical and their
dynamics is responsible for the formation of a
screening mass for the 2D gluon. The loss of its massless
character entails a radical change in a confinement
order parameter such as the v.e.v. of the Polyakov loop.
It becomes non-zero thus indicating the presence
of a deconfining phase transition. It is highly remarkable
that even a very small amount of transverse physics is enough
to produce such a drastic change in the thermodynamics
of pure gauge $QCD_2$.

Although the theory cannot be simple even for very
small values of the strip width -it has to account for
a transition phase-, an easier approach may be taken
in that case. The Fock space of the theory changes
radically depending on what side of the critical interphase
we are describing.
The partition function and all thermodynamic
functions behave very differently below and above
$T_c$ because of that circumstance. The advantage
of the small $\ep$ approach is it allows us to simplify considerably
the dynamics of the physical states in both phases, but, at the same
time, without spoiling the structure of their particular Fock spaces. In
another
words, in the small $\ep$ regime
we always find only color singlet states (glueballs) in the confining
phase, and free gluon particles above $T_c$.
However both glueballs and gluons interact
weakly among them in their respective phases.

In this way we observe that the thermodynamics of the
deconfining phase can be accurately described by a
free gas of transverse and longitudinal gluons. Unquestionably
we cannot predict the appearance of typical non-analyticities
at the transition point with such a simple model.
However the approach is as good as
one wishes (just by manipulating $\ep$) provided we exclude
the singular point. A clear signal that the non-analytical behavior
is really included within the formalism
is encountered in the effective action for the gauge field
modes Eq.(\ref{ActionStructure}). The non-zero modes
are massless. If we integrated them in order to get the
effective potential for the gauge field zero mode
(an order parameter of confinement), we would face strong
IR divergences already at one loop level.
In finite temperature field theory this is a clear indication
of a non-analytical behavior in T.
This non-analyticity
will show up in the form of infrared divergences in the
perturbative expansion
(even off mass-shell).
Typical perturbative
high T calculations break down,
and one has to resort to
complicated  infinite resummations to
find the non-analyticity behavior and predict $T_c$ \cite{Jackiw}.
Fortunately in our case, with the exception of the critical point,
all these effects
are suppressed above $T_c$ when $\ep$ is very small.

The confining phase consists of weakly interacting gluon boundstates,
which at low T behaves as a mix-up of inert glueball gasses.
This fact is directly associated to the NR character of the low-lying
spectrum. However the spectrum also contains fully relativistic
states since there is no upper bound for the binding energy.
These states form the high part of the spectrum. The high
glueball spectrum is extremely populated because it is
made out of many multi-gluon states whose energy
levels are very close to each other. The number
of such states grows fastly with the mass,
as a relativistic treatment points out \cite{dk94}.
They exhibit the Hagedorn behavior.

After heating the system sufficiently we can get to excite
the Hagedorn spectrum. However this supply of heat
does not turn immediately into an increasing of the
gas temperature. Most of the energy is used to excite
the great amount of massive states accessible to the
system instead of increasing the thermal motion of
the less energetical states. The energy necessary to
produce a small change in the temperature becomes
increasingly higher. The gas pressure blows up
at a limiting temperature $T_0(\ep)$ due to the exponential
raise of the density of states. There exists then a temperature
for which the pressure of the glueball gas becomes
higher than that of the free gluon gas. Thermodynamically
the latter is favored and the system evolves to this
phase until the thermodynamical equilibrium is restored.

{}From the microscopical point of view we can understand
the phase transition as produced by a certain instability
arising as a result of increasing fluctuations.
We have seen that multi-gluon states are allowed by
gauge dynamics on the strip. They are
extremely heavy and thus the probability of a multi-gluon
fluctuation is very small at low T. However we know that
the pressure of the glueball gas grows rapidly near the
Hagedorn temperature. The reason for such behavior
is that an increasing amount of energy is needed to excite the large
quantity of multi-gluon states with similar masses,
formed by an increasing number
of constituent gluons (this number grows
with the glueball mass \cite{dk94}). Thus
the probability of generating such multi-gluon fluctuations with
a huge number of constituent gluons becomes
bigger and bigger as we approach $T_0$. It is natural then
to think of $T_c$ as the temperature for which the multi-gluon
fluctuation involves an infinite number of gluons.
The glueball gas turns unstable and the gluon plasma is formed.

As it happened with the glueball spectrum
of the model, the results here presented
show an appealing agreement
with our knowledge of the thermodynamics
of higher dimensional gauge theories. It is
interesting to remark why one expects these
results to describe, at least qualitatively, the features
of the 3D theory.
Our approach consists in defining the whole 3D
theory under special boundary conditions (the
strip compactification). The analysis developed here
and in our previous work \cite{fj95}
accounts for the behavior of pure gauge $QCD_3$
in the regime of very small witdths.
The dynamics of the 2D effective action Eq.(\ref{EffectiveAction}),
obtained to describe the long distance longitudinal
physics of pure gauge $QCD_3$,
is defined completely by the 3D parameters ($g_3$,$L$).
The 2D effective coupling constant and the
2D scalar mass Eq.(\ref{PhiMass}) are fixed by
the 3D inputs. In this sense one deals with the real
theory and not with a 2D model of it. In the latter
we would take $g_2$ and $\mu$ as free parameters
to tune after comparing to 3D data.

Certainly our analysis is restricted to a validity
region ($\ep \! \ll \! 1$) which lies far from the realistic domain,
the infinite width limit. However our unreachable
non-perturbative {\em transverse} regime ($\ep \! \approx \! 1$)
is the natural scenario for lattice simmulations. The study of
asymmetric lattices (one lattice dimension substantially bigger
than the others) complements perfectely our results. The small
$\ep$ regime is precisely the forbidden territory for the lattice simulation,
which is limitated by the finite lattice
spacing constraint. When both analysis merge, one discovers the
two following relevant features \cite{lattice}:

\begin{enumerate}
\item One can extrapolate the small $\ep$
results (i.e., in Eq.(\ref{GlueballMass})) to the $\ep \! \geq \! 1$ regime
in a continuous way. This gives rise to a non-zero value for the $QCD_3$
glueball
masses and the string tension
in the $\ep \! \rightarrow \! \infty$ limit.

\item There exists a stabilization of the $\ep$ flow at a critical $\ep^\ast$
(corresponding to the critical width $L^{\ast}$).
The string tension and the glueball spectrum tend quickly to a constant, their
$QCD_3$ value,
above the critical point $\ep^\ast$.

\end{enumerate}

The previous results strongly suggest that the description of the
low energy longitudinal dynamics of pure gauge $QCD_3$ can
be performed using a generalization of the 2D effective
action Eq.(\ref{EffectiveAction}). That is, through an effective
action defined at the critical $\ep^\ast$,

\beq
\lag_{\mbox{{\footnotesize {\it strip}}}}(\ep^\ast)
         = \mbox{Tr} \left[ \um f_{\alpha\beta}^{\ast 2}
               +  (D_\alpha^\ast \phi)^2 + \m^{\ast 2} \phi^2 +
                        U(\phi; g_2^\ast, \ep^\ast) \right]
\label{RealEffectiveAction}
\eeq

The low energy effective action contains only the terms with less number of
covariant derivatives in the zero mode fields $a_\a$ and $\phi$.
All the parameters defining the 2D effective action get frozen at the
critical width. In particular, the 2D coupling constant and the scalar
mass acquire their infinite width value already at $\ep^\ast$, ($g_2
\!\rightarrow \! g_2^\ast$,
$\mu \! \rightarrow \! \mu^\ast$). The same happens to the effective potential
$U$.

The previous consideration allows us to understand why it is
possible to get a sensible model of pure gauge $QCD_3$
using 2D adjoint matter gauge theories of the type
(\ref{RealEffectiveAction}) \cite{AdjointMatter,ht95,dk94}.
In this approach, the coupling constant $g_2$ and the
scalar mass $\mu$ are no longer fixed by the 3D theory.
They are just input parameters. We cannot know them
until we solve the underlying theory
in the critical region. However it is reasonable to
expect that there exist a set of 2D parameters which
may describe accurately the low energy dynamics
of the 3D theory.

The qualitative
features of the string tension and the glueball spectrum are kept along
their trajectories in $\ep$ space. They get stabilized at the
critical value of the strip width $L^\ast$, which is
identified as the flux tube width. Therefore if the zero
temperature properties of the model
do not seem to disappear in the infinite volume limit,
we have good reasons to believe that the qualitative
picture here developed for the deconfinig phase
transition can be also preserved. With maybe only one
exception, the gluon condensate. Our analysis does
not predict any difference between its values in
both phases.

\section{Acknowledgments}
The authors are specially grateful to V. Vento
for extensive discussions and his strong support. One of us (A.F.) has
benefited from illuminating discussions with D. Espriu, J. Soto, J. Taron
and A. Travesset.

\newpage
\begin{center}
\large{Figure Captions}
\end{center}

Fig.1: Diagrams contributing to the $a_n$ self-energy at high T.
Dashed lines correspond to $\phi$ modes, wavy
lines correspond to gluon modes.

Fig.2 : Pressure versus temperature in the relativistic confining (dashed
lines),
NR confining (dot-dashed lines) and deconfining (solid lines)
phase ($g_3 \!=\! 1,\ep \!=\! 0.01$). The vacuum pressure is taken to be
zero in both phases (see text): (a) The critical crossing for
a relativistic high energy spectrum; (b) The critical crossing for
the NR spectrum.

Fig3: Estimated evolution of the critical temperature with the strip width ($L
\!=\! \ep$,
$g_3 \!=\! 1$).

\end{document}